\documentclass[epj,referee]{svjour}

\usepackage{graphicx}
\usepackage{latexsym}

\title{Complex Grid Computing}

\author{
  Luciano da Fontoura Costa \and 
  Gonzalo Travieso \and 
  Carlos Antonio Ruggiero
}            

\institute{
  Instituto de F\'{\i}sica de S\~{a}o Carlos,
  Universidade de S\~{a}o Paulo, Caixa Postal 369, 13560-970, S\~{a}o
  Carlos, SP, Brazil,\\
  \email{luciano@ifsc.usp.br,gonzalo@ifsc.usp.br,toto@ifsc.usp.br}
}

\date{Received: date / Revised version: date}

\abstract{
  This article investigates the functional properties of complex
  networks used as grid computing systems. Complex networks following
  the Erd\H{o}s-R\'{e}nyi model and other models with a preferential
  attachment rule (with and without growth) or priority to the
  connection of isolated nodes are studied. Regular networks are also
  considered for comparison. The processing load of the parallel
  program executed on the grid is assigned to the nodes on demand, and
  the efficiency of the overall computation is quantified in terms of
  the parallel speedup. It is found that networks with preferential
  attachment allow lower computing efficiency than networks with
  uniform link attachment. At the same time, considering only node
  clusters of the same size, preferential attachment networks display
  better efficiencies. The regular networks, on the other hand,
  display a poor efficiency, due to their implied larger internode
  distances. A correlation is observed between the topological
  properties of the network, specialy average cluster size, and their
  respective computing efficiency.
  \PACS{
    {89.75.Fb}{Structures and organization in complex systems} \and
    {89.20.Ff}{Computer science and technology} \and 
    {89.20.Hh}{World Wide Web, Internet} \and 
    {02.10.Ox}{Combinatorics; graph theory} \and 
    {89.75.Hc}{Networks and genealogical trees} 
  } 
}

\authorrunning{Costa, Travieso and Ruggiero}

\begin{document}

\maketitle

\section{Introduction} 
 
Among the many implications of the scientific and technological
advances in microelectronics along the last decades, the availability
of microprocessors characterized by ever diminishing size, cost and
power consumption (per operation), together with increasing computing
power \cite{Culler99}, has led to an unprecedented opportunity for
parallel computing, allowing simulations of non-linear and complex
physical systems.  More recently, the advent of the Internet paved the
way for using a network of computers to obtain a very large and
powerful computing system, defining the new research areas of
\emph{grid computing} \cite{Skillicorn01,Foster01,Fox01} and
\emph{parasitic computing} \cite{Barabasi01}.

A parallel computing system consists of a set of processing elements
connected by some kind of communication network. A parallel program
runs on the system by partitioning the work to be done in several
pieces that are executed on the available processing elements.  To
collaborate in the execution of the program, each piece must, as a
rule, communicate with others through the communication network. To
show a good performance in the execution of a program, a parallel
system must then: (i)~make a large number of processing elements
available to the application; and (ii)~enable fast communications
between these processing elements.

For grid computing, the first condition can be met by the large
amount of computers available in the Internet, as long as their
owners agree to make their processing power available. But the second
condition is difficult to meet: even if high bandwidth networks are
common, the latency to deliver a message to a site geographically far
from the origin is large. The increased processing power of
microprocessors due to Moore's law (i.e.\ the number of transistors in
a chip doubles each 1.5~years) contrasts with the typically slow
interconnection between processing elements, thus undermining the
performance of parallel systems due to the relatively large time spent
to send data for processing elsewhere in comparison with the time
taken to process that data. A program will only have good performance
on the grid if the amount of computations done is large in comparison
with the time needed to send a message between two processors. Thus,
the effective use of grid computing remains a challenge, demanding a
delicate balance between computation and interprocess communication
workloads.  As a rule, the weaker the coupling (i.e.\ the amount of
communication needed) between different processing tasks, the higher
the overall efficiency in a given parallel system.

Generally, a more densely connected processing network favors faster
data transmission, as the mean distance between nodes tends to
decrease, but at the expense of additional communication resources.
Moreover, the specific kind of processing to be performed and the
availability of the computer resources for collaborative (or
parasitic) computing also play an important role in defining the
overall grid execution performance.

The novel area of complex networks (e.g.\ 
\cite{Albert02,Dorogovtsev02,Newman03}) has drawn increasing interest
of the physics community.  A large part of the success of such an
approach derives from the fact that such networks have been found to
adhere, at varying degrees, with important real phenomena such as
transmission of infectious diseases, social and ecological interactions,
and the Internet.  The fact that today most computers are interconnected
through the Internet has contributed further to promote the systematic
investigation of the Internet characteristics, a task that can greatly
benefit from physical modeling approaches.
 
As several features are shared by grid computing systems and complex
networks, much can be gained through integrative and comparative
approaches, allowing cross-fertilization between those two important
areas.  The underlying idea in the current work is to study the
efficiency of parallel/distributed architectures whose
interconnections are defined in terms of complex network models.  Such
an investigation therefore focus on the integration between topology
and function of the networks, an important aspect of complex network
research \cite{Newman03}. Regular networks, which are often used in
parallel computing, are also considered as a reference for comparison.
It is particularly interesting to verify how the specific properties
of these interconnecting schemes ---such as the average vertex-vertex
distance and cluster size--- affect the processing time and efficiency
for different network configurations.  Such a possible dependence
between the topological and functional properties of the networks is
backed by recent works which verified that the emergent features of
complex networks, such as associative memory recall in neuronal
networks, can be strongly affected by the network interconnecting
scheme and phase transitions \cite{Stauffer03,Costa03}.
In related works, the complex network paradigm has also been explored
from the perspective of search algorithms
\cite{Kleinberg00,Adamic02,Guimera02} and information transfer in graphs
\cite{Sole01,Tadic04,Tadic05}.

\section{Network Models Used}

The main purpose of this article is to study the influence of some
network topological features in the efficiency of a grid system for
the execution of a suitable parallel program. Therefore, we will not
try to use network models that reproduce the characteristics of the
Internet; we instead restrict ourselves to some simple models which
are described below. The models are undirected, reflecting the
bidirectional transfer of packets on the Internet.

Let a complex network be represented as a graph with $n$ nodes,
identified as $i$, $i=0, \ldots, n-1$, and unweighted,
undirected edges represented as $(i,j)$. The first model
considered is the Erd\H{o}s-R\'enyi (ER) model with a fixed number $c$
of edges. In this model, for each connection, two nodes $i$ and $j$ are
chosen uniformly among all the nodes to establish the connection
$(i,j)$. Self-connections (connections of a node with itself) and
duplicate connections (connections between already connected nodes) are
avoided in this and all the following models.

ER graphs have a fast decaying degree distribution, with very small
probability for nodes with high degree (also called \emph{hubs}). To
widen the degree distribution and increase the probability of high
degree nodes, a \emph{preferential attachment} (PA) model is
used. Networks in this model are generated as described in
\cite{Stauffer03}: starting with all nodes without connections and
choosing two nodes to connect by drawing nodes from a list of node
numbers represented in amount proportional to their respective number
of connections (plus one, to account for the unconnected nodes). Note
that this network model, although having preferential attachment, has
no growth and so does not lead to scale-free networks (see
Section~\ref{nettop}).

Hubs of larger degrees can be found in scale-free networks. A simple
model for scale-free networks was proposed by Barab\'asi and Albert
\cite{Barabasi99}. In their model, the network starts with $m_0$
connected nodes and grows by the addition of one node at a time. When a
node is added, $m$ new connections from the new node to already existing
nodes are made, and each already existing node can be chosen to receive
a connection with probability proportional to its degree $k$. In this
model, all nodes in the network form a single large connected component;
as we are interested in studying the influence of percolation with
growing connectivity, their model is not adequate due to the
nonexistence of a percolation transition and to the impossibility of
specifying an average connectivity that is not an even integer (the
average degree is always $2m$). For this reason, their model is here
generalized as described in the following. In the model used in this
work, which we call \emph{scale-free} (SF) model, instead of having a
fixed number of connections for each new node $i$, a random number $m_i$
is chosen using a Poisson distribution with mean $m$, and $m_i$
connections from this node to the already existing nodes are made. As
some nodes may have $m_i=0$, the network will have unconnected nodes; to
enable these nodes to receive connections with the addition of new
nodes, each already existing node is chosen with probability
proportional do $k+1$ instead of $k$. In this model, as $m$ is only a
mean value, it can be any real number (instead of only an integer
number, as in the Barab\'asi-Albert model); also, as new nodes are not
necessarily connected to the already existing nodes, the network
consists of many connected components.

For grid computing, as for all kinds of collaborative work between the
agents represented by a network, the binding of the agents to other
agents of the network is a necessity. There is, therefore, a tendency
to bind nodes to the network as new communication resources are
available, instead of using them to bind already connected nodes. This
suggests a different kind of random network construction: connecting
new nodes to the network should be given preference while there are
still isolated nodes. For that purpose, we introduce here two new models
of random networks. The parameters for their construction are: the
number of nodes $n$ and the number of connections $c$.

In the first model, one end of each new edge is chosen with uniform
distribution among the isolated nodes and the other end with uniform
distribution among all nodes (isolated or not). If at some point no
more isolated nodes exist, both ends of the remaining edges are drawn
with uniform distribution among all nodes. 

In the second model, one end of the new edge is drawn randomly from
the set of isolated nodes, as for the previous model, but the other
end is drawn among all nodes with probability proportional to the node
degrees, as in the previously discussed preferential attachment
network. 

We call these networks \emph{insertion networks}, because the
connections are used to insert the nodes in the network; the first
model is called \emph{uniform-uniform} (UU) insertion network, and the
second model \emph{uniform-preferential} (UP) insertion network.
 
For comparison, we study also three regular network structures common
in parallel systems \cite{Stojmenovic96}: the hypercube and the 2D
and 3D tori. The construction of an hypercube with $n$ nodes can be
explained as follows. For node $i$, represent the value of $i$ in
binary using $\lceil\log_2n\rceil$ bits; there is a link $(i,j)$ iff the
binary representation of $j$ differ in exactly one bit from that of $i$.
For example, in an $n=8$ network, node 3 (binary 011) is connected to
nodes 2 (010), 1 (001) and 7 (111).  In the bi-dimensional torus the
nodes are distributed in a grid of size $(n_x, n_y)$ (with $n_x n_y =
n$), each node receiving a label $(x,y)$, where $x = \lfloor
i/n_y\rfloor$ and $y = i \bmod n_y$; node $(x,y)$ is connected with node
$(x',y')$ iff $x'=x$ and $(y' = y\pm1) \bmod {n_y}$ or $y'=y$ and $(x'
=
x\pm1) \bmod {n_x}$. For example, in a $n=8$ bi-dimensional torus
organized as a $2\times4$ grid, node 3 corresponds to coordinates
$(0,3)$ and is connected to nodes 2 $(0,2)$, 0 $(0,0)$, and 7 $(1,3)$.
The three-dimensional torus is similarly constructed.
 
\section{Results}

The results are divided in two parts. First, some results concerning
the properties of the network models described are presented. Then,
results obtained by using these network models as communication
infrastructure for the simulation of the execution of a parallel
program on a grid are shown.

\subsection{Network properties}\label{nettop}

A thorough analysis of the network models described is out of
the scope of this paper. Here only some properties of interest to the
analysis of the following grid simulation results
(Section~\ref{gridsimul}) are presented.

The topological properties of those networks are quantified in terms
of the following measures: (a)~node degree $k$; (b)~mean vertex-vertex
distance $\ell=\frac{2}{n(n-1)}\sum_{i>j}d_{ij}$ where $d_{ij}$ is the
geodesic distance (distance, in number of links, of the shortest path)
between nodes $i$ and $j$ and the summation includes only pairs $ij$
that have a path connecting them; and (c)~mean cluster size
$s$. A cluster, also know as a connected component, is a set of directly
or indirectly connected nodes, i.e., nodes that can be reached from all
the other nodes on the cluster by a path. To evaluate the mean cluster
size, we compute for each node of the network the number of nodes on its
cluster and take the average of these values for all nodes on the
network. Note that this average includes the largest cluster, and is
dominated by it after percolation. To enable the comparison between the
different complex networks models with respect to their connectivity, we
use the parameter $z$ defined as $z = 2m$ for the SF model and $z =
2c/n$ for the other models. In the limit of large $n$ we have $z =
\left<k\right>$ for all network models.
 
In the following, network characteristics are quantified as averages
for 50~random networks for each set of parameters. Error bars display
the 99\% confidency interval for the computed average, considering
normal distribution of the averaged values.

\begin{figure*}
  
  \hspace{0pt plus 1fil}
  \includegraphics[width=0.22\textwidth]{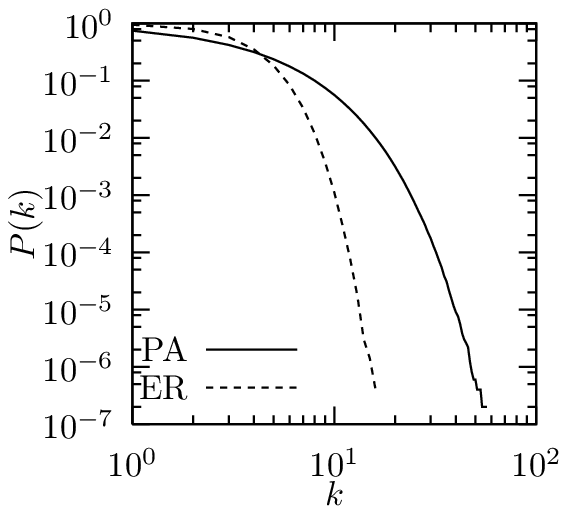}
  \hspace{0pt plus 1fil}
  \includegraphics[width=0.22\textwidth]{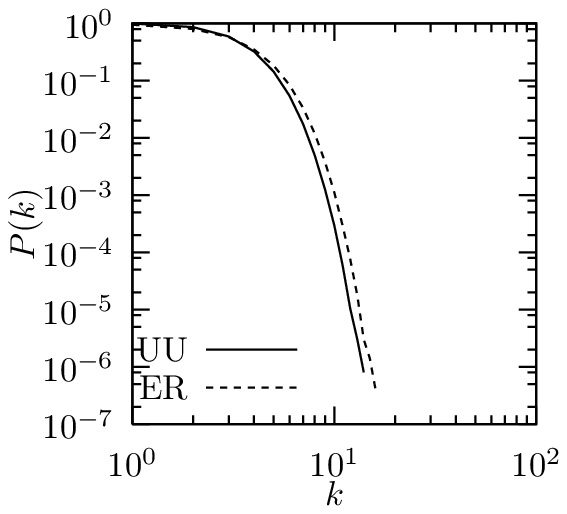}
  \hspace{0pt plus 1fil}
  \includegraphics[width=0.22\textwidth]{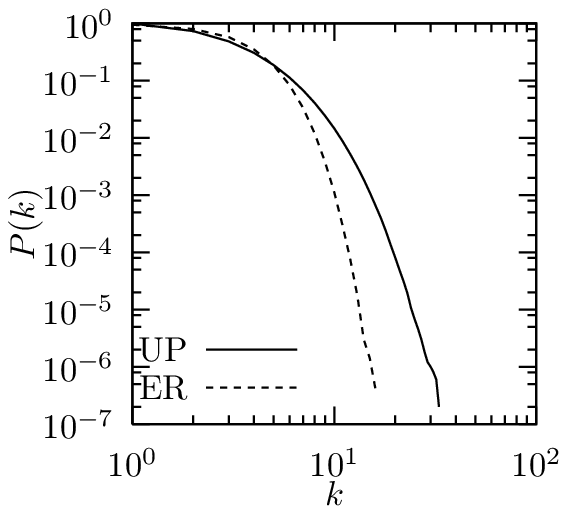}
  \hspace{0pt plus 1fil}
  \includegraphics[width=0.22\textwidth]{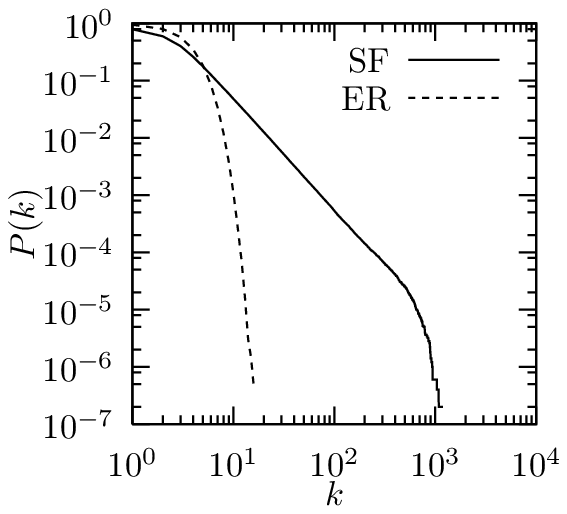}
  \hspace{0pt plus 1fil}
  \\
  \centerline{\hspace*{1cm}(a) \hspace{3.8cm} (b) \hspace{3.8cm} (c)
    \hspace{3.8cm} (d)\hspace*{1cm} }

  \caption{Degree distributions for the network models. The ER 
    model results appear in all figures, as a reference for comparison.
    The networks have $n=100000$ nodes and $z=3$: (a) preferential
    attachment; (b) insertion with uniform probability; (b) insertion
    with preferential attachment; and (d) scale-free.}

  \label{fig:degree}

\end{figure*}

The degree distributions for each of the considered models for
$n=100000$ and $z=3$, compared with the ER model, are presented in
Figure~\ref{fig:degree}, where $P(k)$ is the cumulative probability
distribution (probability of finding a node in the network with degree
larger than $k$). The UU model is almost indistinguishable from the ER
model in terms of degree distribution. Due to the preferential
attachment rule, the PA and UP models have broader degree
distributions. For the UP model this effect is not so marked due to
the preference given to newly added nodes in new connections. As
expected, the SF model follows a power law (with a finite-size
cutoff), given rise to a large probability of nodes with high degree.
 
\begin{figure*}

  \hspace{0.6cm plus 1fil}
  \includegraphics[width=0.6\columnwidth]{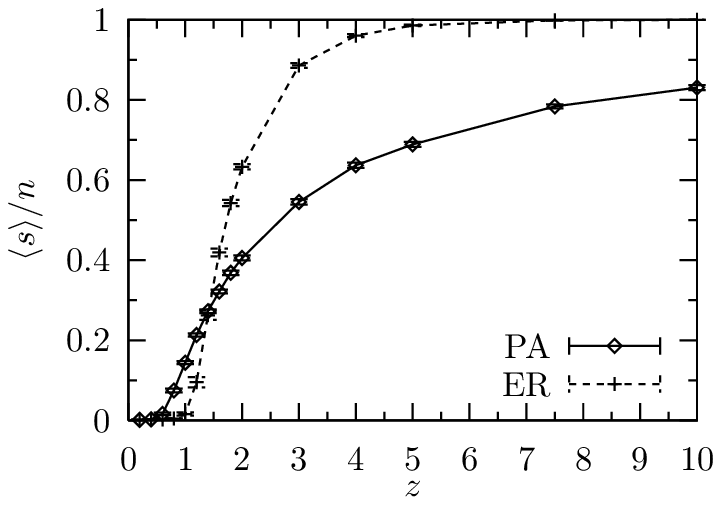}
  \hspace{0pt plus 1fil}
  \includegraphics[width=0.6\columnwidth]{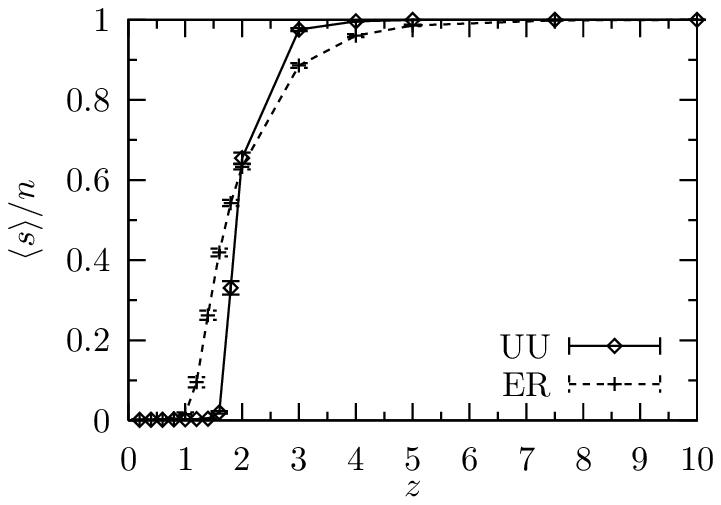}
  \hspace{0pt plus 1fil}
  \\
  \centerline{(a) \hspace{6.5cm} (b) } \\ \\
  \hspace*{0.6cm plus 1fil}
  \includegraphics[width=0.6\columnwidth]{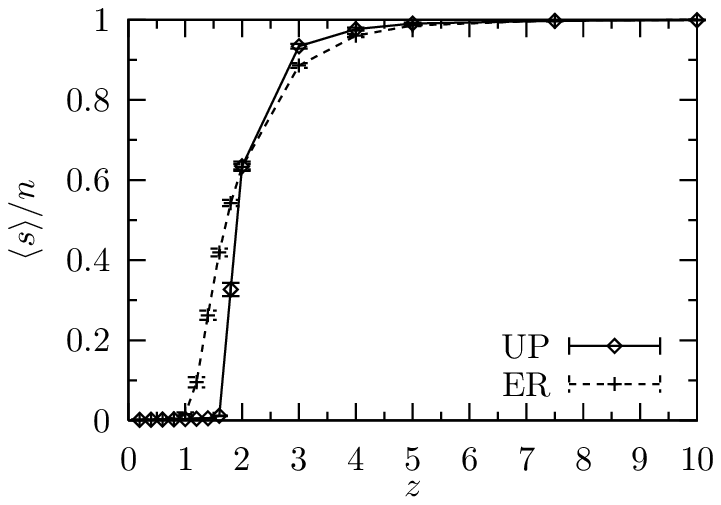}
  \hspace{0pt plus 1fil}
  \includegraphics[width=0.6\columnwidth]{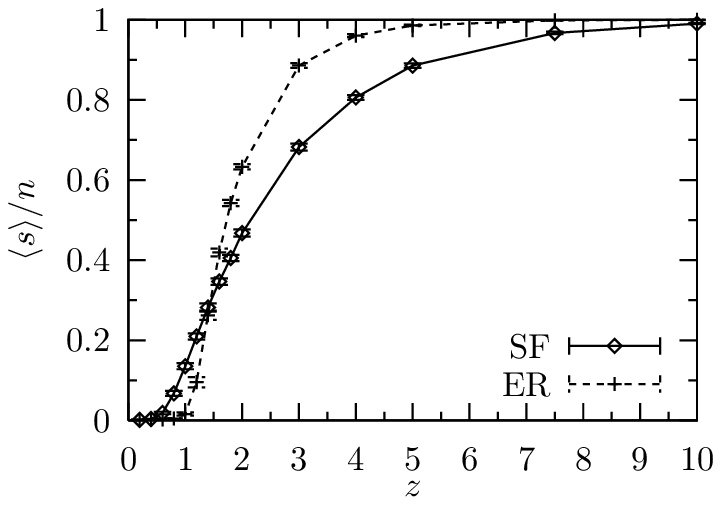}
  \hspace{0pt plus 1fil}
  \\
  \centerline{(c) \hspace{6.5cm} (d) }

  \caption{Normalized average cluster size, for $n=1000$, for the PA
    (a), UU (b), UP (c) and SF (d) models, compared with the ER model.}

  \label{fig:cs}

\end{figure*}

\begin{figure*}

  \hspace{0.6cm plus 1fil}
  \includegraphics[width=0.6\columnwidth]{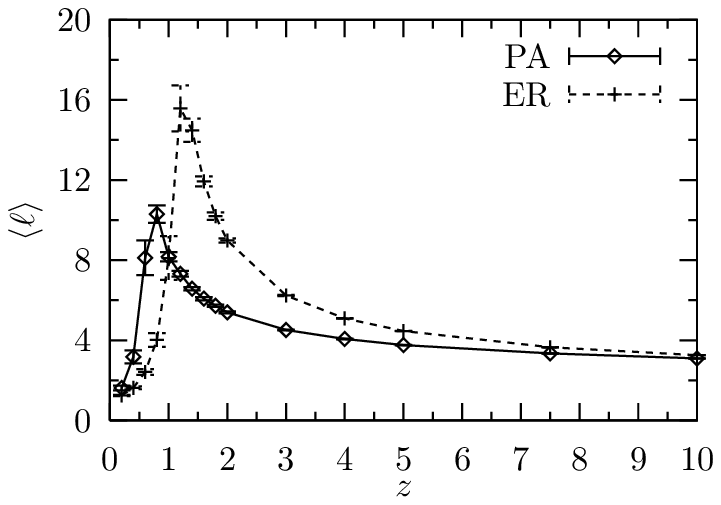}
  \hspace{0pt plus 1fil}
  \includegraphics[width=0.6\columnwidth]{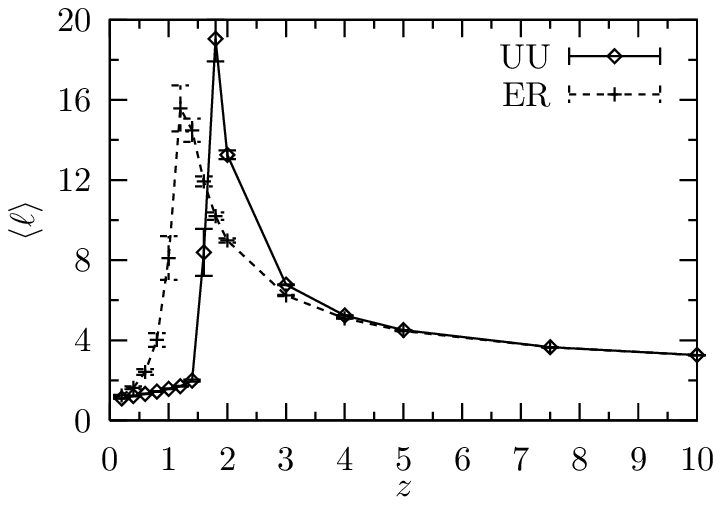}  
  \hspace{0pt plus 1fil}
  \\
  \centerline{(a) \hspace{6.5cm} (b) } \\ \\
  \hspace*{0.6cm plus 1fil}
  \includegraphics[width=0.6\columnwidth]{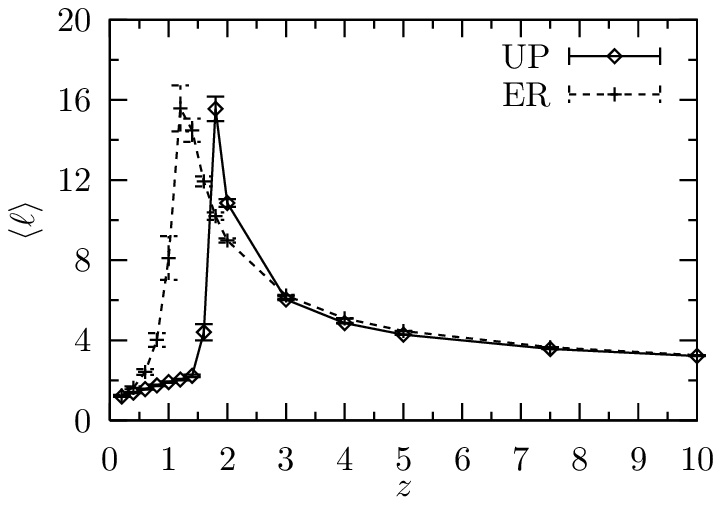}
  \hspace{0pt plus 1fil}
  \includegraphics[width=0.6\columnwidth]{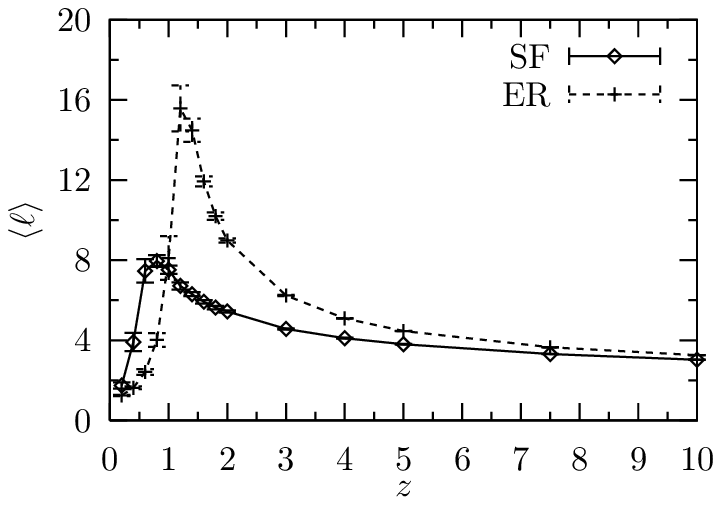}
  \hspace{0pt plus 1fil}
  \\
  \centerline{(c) \hspace{6.5cm} (d) }
 
  \caption{Average distance between connected pairs, for $n=1000$, for
    the PA (a), UU (b), UP (c) and SF (d) models, compared with the ER
    model.}

  \label{fig:l}

\end{figure*}

Fig.~\ref{fig:cs} gives the average network cluster size $\left< s
\right>$ normalized by the network size $n$, and Fig.~\ref{fig:l} shows
the average node distance $\left<\ell\right>$ for the networks.  These
results are shown as functions of $z$ comparatively to the
ER model.  As Figs.~\ref{fig:cs}(a--d) show, there is an abrupt
transition (a percolation transition) from small to large cluster sizes
as the connectivity grows. In the small cluster size region, the mean
distance tends to grow with the connectivity (Fig.~\ref{fig:l}), as new
links result in new connections between previously unconnected nodes; in
the large cluster region, increased connectivity reduces mean distance,
as most of the nodes are already connected in the largest cluster. A
striking feature of the results for the PA model (and to some extent
for the SF model) is the small cluster sizes even for high connectivity.
The reason is that, as the number of nodes connected in the largest
cluster grows, the probability of linking an unconnected node is very
small, due to the preferential attachment rule used to choose the ends
of new links, resulting in a relatively large number of isolated nodes
or small clusters. In the SF model, this problem is minimized by the
fact that one end of each new connection always go to a new
(previously unconnected) node. The insertion (UU and UP) models show
very similar behavior in terms of average distance and cluster size. The
formation of a cluster spanning most of the nodes occurs for these
models for higher connectivities than for the ER model. The explanation
is that links that could be used to connect small clusters to form a
larger one are being used to link new nodes. On the other hand, the size
of the resulting largest cluster tends to be larger, as isolated nodes
are less likely to appear. Distances in these models tend to be higher
than in the ER model, because the same number of links is used to
connect a larger number of nodes. For sufficiently higher
connectivities, this last property is compensated in the UP model by the
onset of high degree nodes (hubs), that shorten the mean distance
between the nodes of the cluster. The presence of hubs is also the
explanation for the smaller distances shown by the PA and SF models;
this advantage vanishes as the connectivity grows.

\begin{figure*}

  \includegraphics[width=0.45\textwidth]{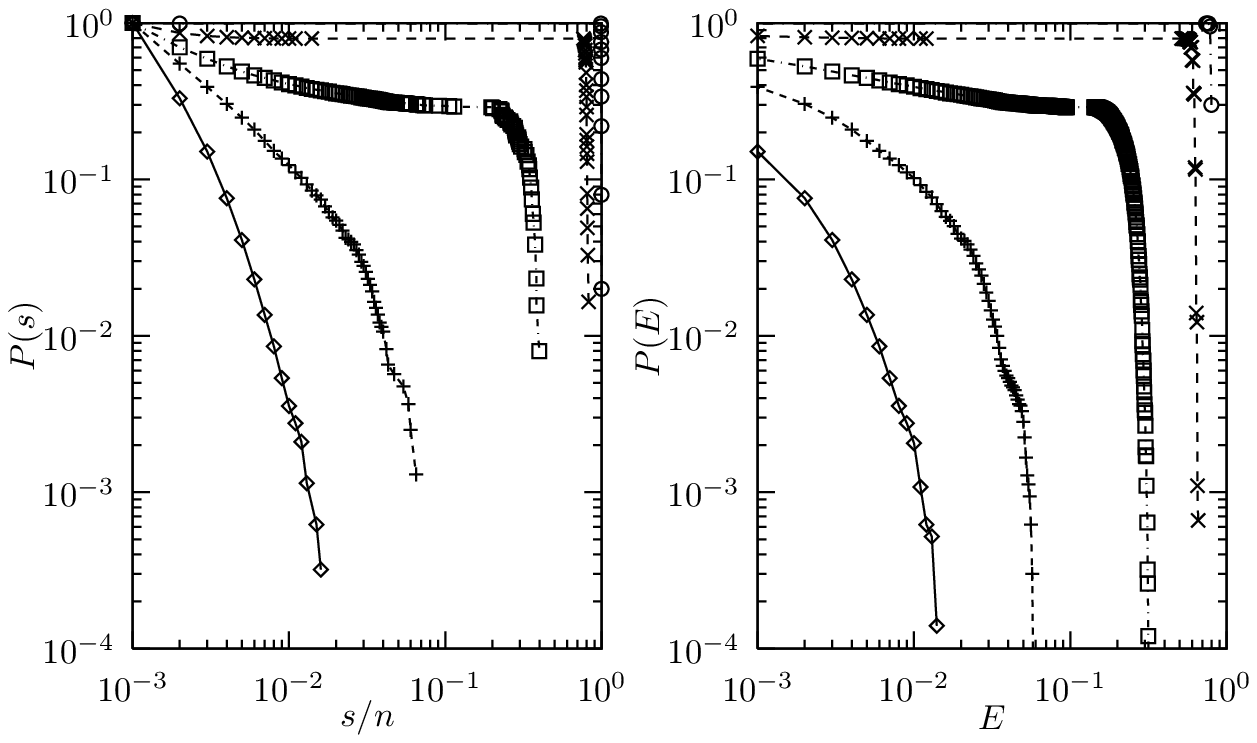}
  \hspace{0pt plus 1fil}
  \includegraphics[width=0.45\textwidth]{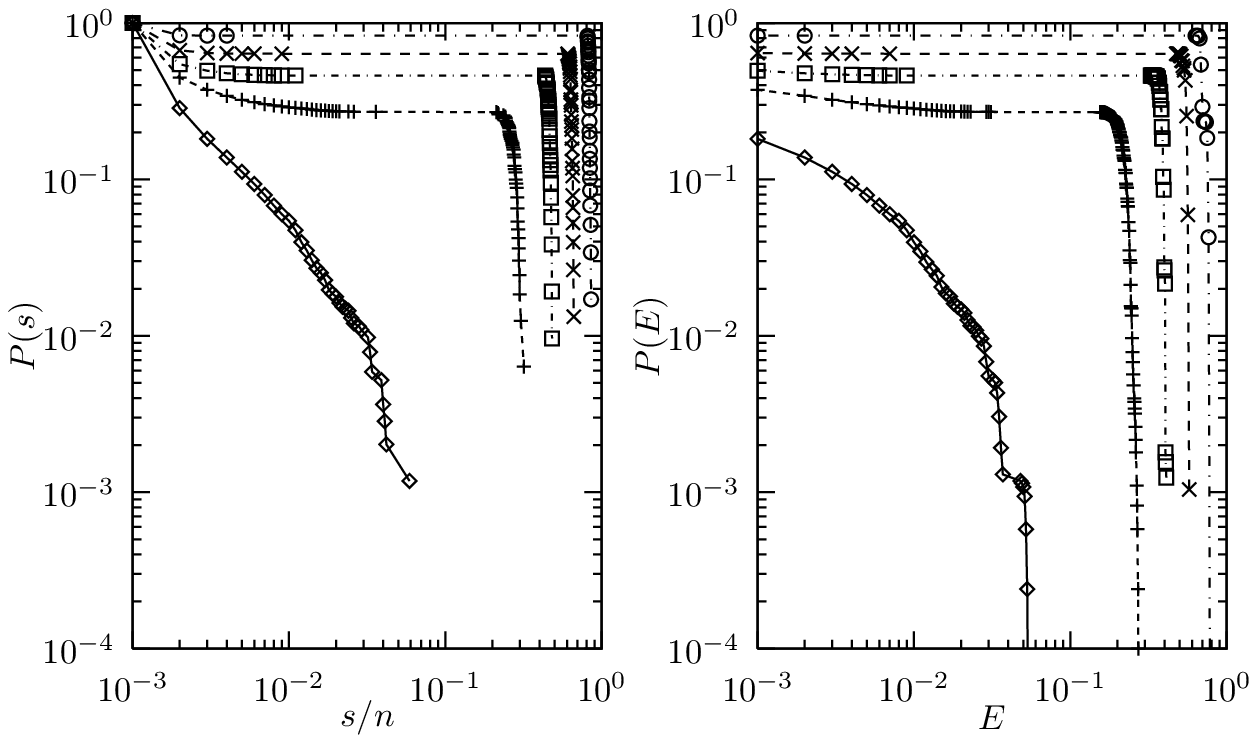} \\
  \centerline{(a) \hspace{8.5cm} (b)}\\ \\
  \includegraphics[width=0.45\textwidth]{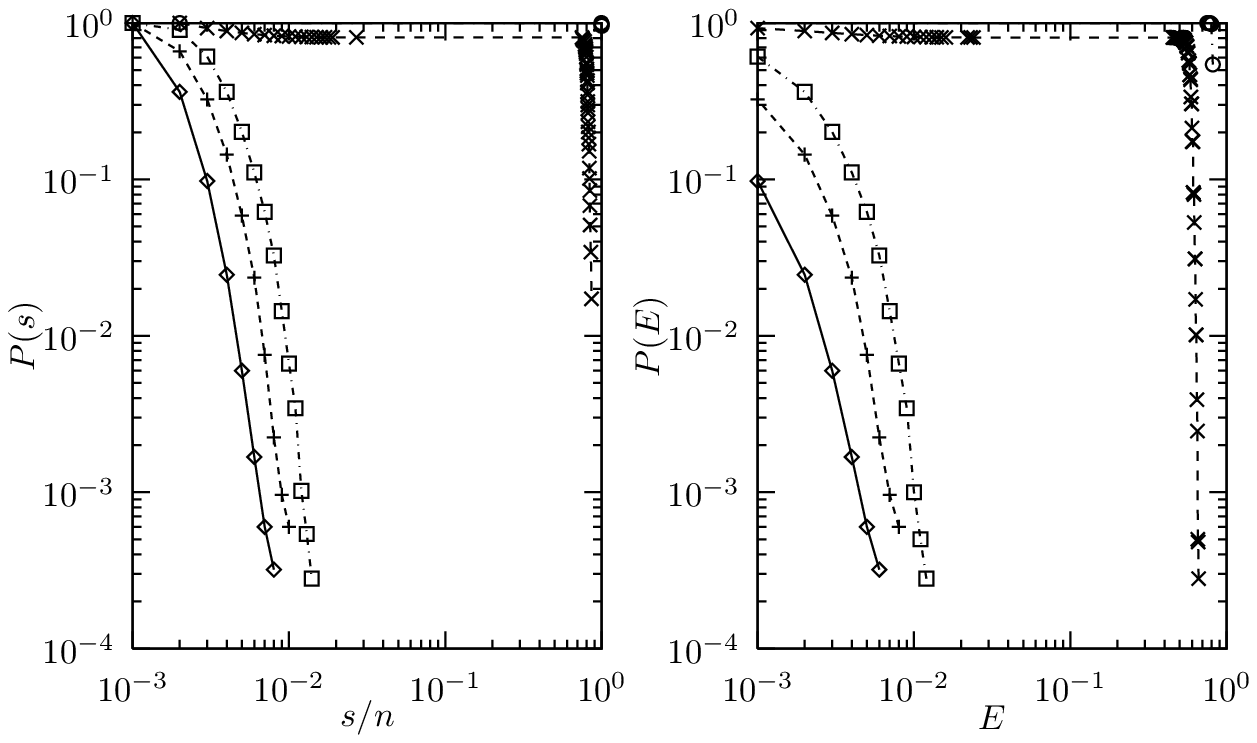}
  \hspace{0pt plus 1fil}
  \includegraphics[width=0.45\textwidth]{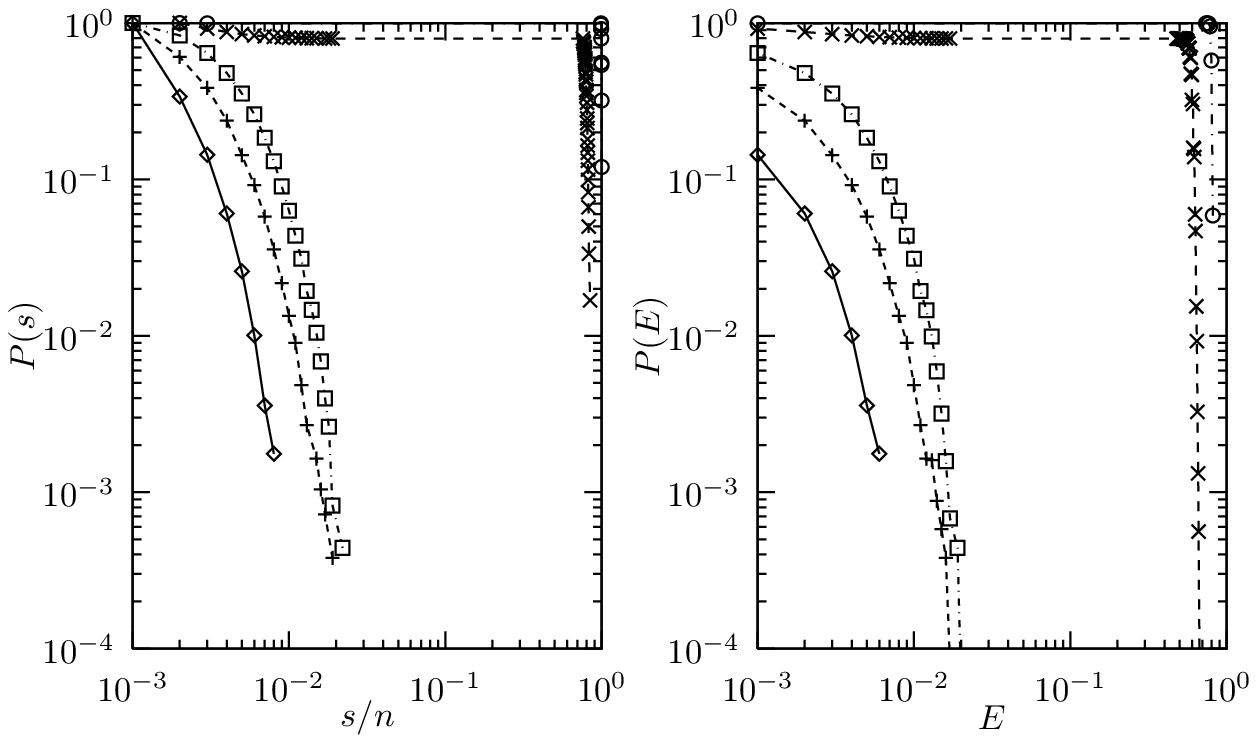}\\
  \centerline{(c) \hspace{8.5cm} (d)}\\ \\
  \centerline{\includegraphics[width=0.45\textwidth]{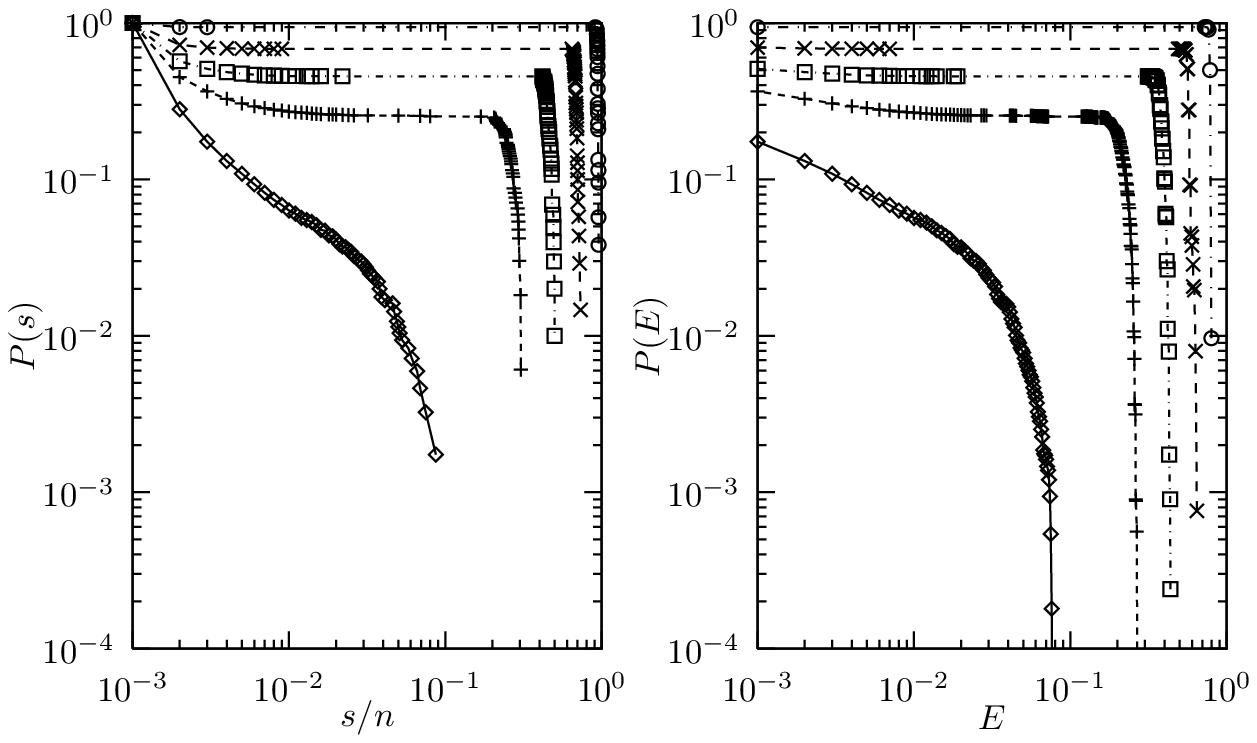}}
  \centerline{(e)}
  
  \caption{Cumulative probability distribution of cluster size $P(s)$
   (probability of a randomly selected node being part of a cluster of
   size $s' \ge s$) in terms of $s/n$ and parallel efficiency $P(E)$
   (probability of a randomly selected node achieving efficiency $E'
   \ge E$ when chosen as master) for the five models: (a)~ER; (b)~PA;
   (c)~UU; (d)~UP; and (e)~SF. Distributions shown for five different
   average connectivity ($z$) values: $z=0.4$ ($\Diamond$); $z=0.8$
   ($+$); $z=1.2$ ($\Box$); $z=2.0$ ($\times$); and $z=5.0$
   ($\circ$). Network size: $n=1000;$ number of work packets
   $M=5000;$ work packet size $L=100.$}
   
  \label{fig:hcs}

\end{figure*}

More detail about the size of clusters is given in the left
side of Figs.~\ref{fig:hcs}(a--e), which display the cumulative
probability distribution of cluster sizes (i.e.\ the probability $P(s)$
that a randomly selected node is part of a cluster of size greater or
equal to $s$) against $s/n$ for some values of $z$. These figures show
that for small average connectivities the probability of finding a large
cluster is negligible. On the other hand, for sufficiently large average
connectivities almost all nodes are found in large clusters. The models
without preferential attachment (ER, Fig.~\ref{fig:hcs}(a), and
specially the insertion models, UU Figs.~\ref{fig:hcs}(c) and UP
Fig.~\ref{fig:hcs}(d)) show a sharp transition from a regime with low
probability of large clusters and high probability of small clusters to
a regime with high probability of large clusters and small probability
of small clusters. On the preferential attachment models (PA,
Fig.~\ref{fig:hcs}(b), and SF, Fig.~\ref{fig:hcs}(e)), this transition
is more gradual; they also display a larger probability of medium
sized clusters before the percolation.

In the regular networks, all nodes are connected, and so $\left< s
\right>/n = 1$. Also, the value of $z$ is fixed for each network type,
given $n$. The third column of Table~\ref{tab:reg} shows the values of
the average distance for these network types.

\begin{table}

  \caption{The topological and efficiency measurements for the three
    considered regular topologies ($n=1000$).}
    
  \label{tab:reg}
  
  \begin{center}
    \begin{tabular}{rccc} 
      \hline\noalign{\smallskip}
      Network type  &  $z$  & $\left<\ell\right>$ & $\left< E \right>$ \\ 
      \noalign{\smallskip}\hline\noalign{\smallskip}
      2D torus      &    $4$          &   $16 \pm 7$ &    $0.664$   \\
      3D torus      &    $6$          &   $7 \pm 3$  &    $0.778$   \\
      Hypercube     & $9.9 \pm 0.4$   &   $5 \pm 2$  &    $0.816$   \\
      \noalign{\smallskip}\hline
    \end{tabular}
  \end{center}
  
\end{table}
 
\subsection{Grid Simulations}\label{gridsimul}

The parallel computing systems were obtained by assigning a processing
unity to each network node, while messages flow along the network
edges.  The distributed application considered follows the
\emph{master/slave} paradigm (also known as \emph{manager/worker} or
\emph{bag of tasks}), where a master delivers processing tasks on demand
to slave computers. The computational tasks are assumed to be completely
independent, in the sense that each node can proceed without additional
communications after receiving the work packet. This arrangement is
similar to many grid computing efforts, like SETI@home
\cite{Anderson02}. The computations are partitioned into $M$ work
packets (tasks), each requiring the same amount $L$ of computing time.
The communication cost is taken to correspond to the minimum number of
edges between the master and the slave requesting the data. \emph{The
edge communication overhead is therefore equal for all edges and adopted
as time unit.} Taking into account this communication cost model, the
very small number of short cycles present in all the considered network
models (compared with the Internet) does not represent an additional
limitation.
 
Given a network, each node $i$ at a time is considered as master. The
nodes that are part of the same cluster as the master start requesting
tasks. The nodes that are not part of the same cluster as the master
cannot contribute to the computation because of the lack of connection
to the master. After receiving a task from the master, the slave
computes the result, taking time $L$, and sends it, together with a
request for another task, to the master.  When all $M$ tasks have been
delivered and their results received, the master terminates the
execution and computes its total execution time $T_i$. Isolated nodes
cannot take part on a distributed computation and so, when chosen as
master nodes, their execution time is considered infinite (they will
wait forever to receive a work request from a slave).

To quantify the suitability of the network models for grid computing we
compute the average speedup achieved by the execution of the application
on the networks. The speedup is defined as the ratio between
sequential and parallel execution times. For the problem 
considered, the parallel execution time for master node $i$ is the
value of $T_i$ discussed in the previous paragraph and the sequential
execution time is $ML$, so that $S_i = \frac{ML}{T_i}$ is the speedup.
The mean speedup of a network is the mean value $\frac{1}{n}\sum S_i$.
Note that, for isolated nodes, as discussed, $T_i = \infty$ and so
$S_i = 0$; as these nodes are anyway considered in the average, if a
network has many isolated nodes its average speedup is low.  Another
equivalent measure, used in our results, is the normalized speedup $E
= S/n$, also known as \emph{parallel efficiency}.  The averages
$\left<S\right>$ and $\left<E\right>$ are then taken for 50~different
random networks for each model and parameter set.

If master $i$ sends a task to slave $j$, the time to complete the task
(get the result back) is the sum of the computation time of the task
with the time taken to send the task to the slave node and receive it
back, that is, $L + 2d_{ij}$.  At the same time, the number of nodes
computing tasks is $s_i-1$ ($s_i$ is the size of the cluster of master
node $i$ and we subtract one because the master does not compute
tasks).  This indicates that, for the problem considered, two
important network metrics are the average cluster size $\left< s
\right>$ (Fig.~\ref{fig:cs}) and average distance $\ell$ between nodes
with a path connecting them (Fig.~\ref{fig:l}) .

\begin{figure}

  \begin{center}  
  \includegraphics[width=0.6\columnwidth]{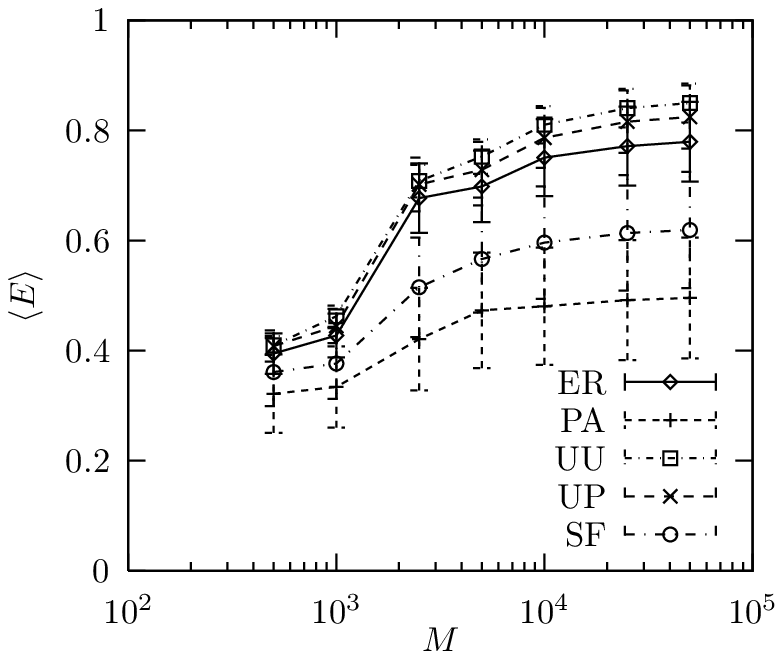} \\
  (a) \\
  \includegraphics[width=0.6\columnwidth]{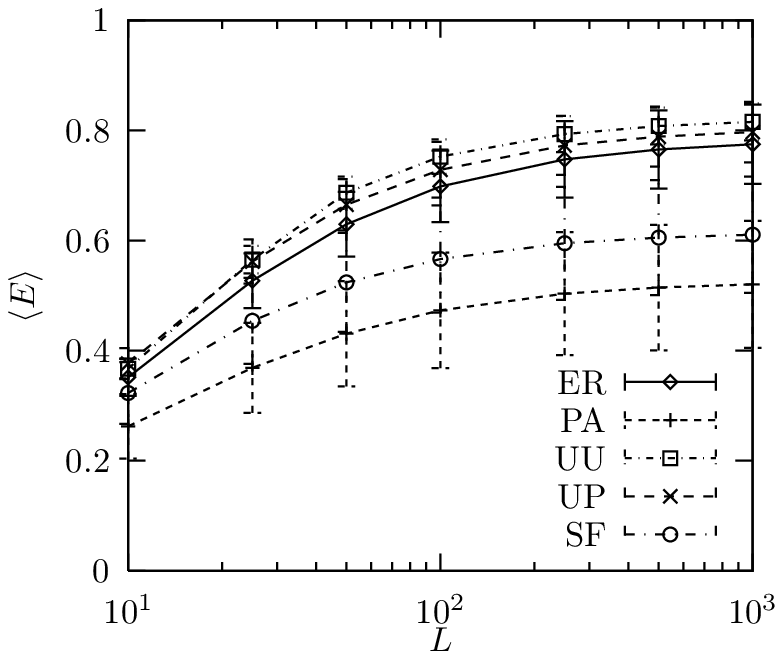} \\
  (b)
  \end{center}
  
  \caption{Parallel efficiency $\langle E\rangle$ as a function of the
    number $M$ (a) and size $L$ (b) of work packets. Network parameters
    are $n=1000$ and $z=3.$ In (a) $L=100$ and in (b) $M=5000.$}

  \label{fig:ml}

\end{figure}

The number of tasks $M$ was chosen so that each node has some tasks to
process, but avoiding too large number of tasks, because the simulation
time is proportional to $M$; the size of the tasks $L$ was chosen to
achieve computation times larger than the average communication time
(two times the average distance; see also Figs.~\ref{fig:l}(a--d)), but
small enough such that the communication time has a detectable effect.
If $L$ is too large, the average communication time, and then the
network structure, is not important; if it is too small, the grid system
is not a good choice for the execution of the application.
Fig.~\ref{fig:ml}(a) shows the dependence of average efficiency
$\langle E \rangle$ with the number of tasks $M$ (for fixed $L=100,$
$n=1000,$ and $z=3$) for the five models. If $M$ is small, there will be
not enough computational work to distribute evenly among the nodes,
resulting in very small efficiencies; after some value of $M,$ the 
efficiency is not much affected by a further increase of $M.$ The
relation of the efficiency with $L$ is very similar: a larger value o
$L$ helps reduce the importance of the communication costs, increasing
the efficiency; Fig.~\ref{fig:ml}(b) plots this relation (for $M=5000,$
$n=1000,$ and $z=3$). It can be seen that after $L\approx 100$ there is
no significant increase in efficiency for larger $L.$ The following
results assume $M=5000$ and $L=100;$ results for other values show no
qualitative differences.

\begin{figure*} 

  \hspace{0.6cm plus 1fil}
  \includegraphics[width=0.6\columnwidth]{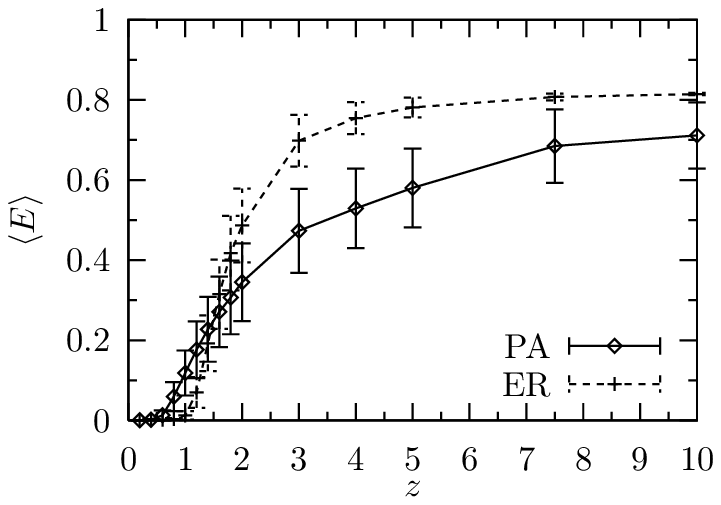}
  \hspace{0pt plus 1fil}
  \includegraphics[width=0.6\columnwidth]{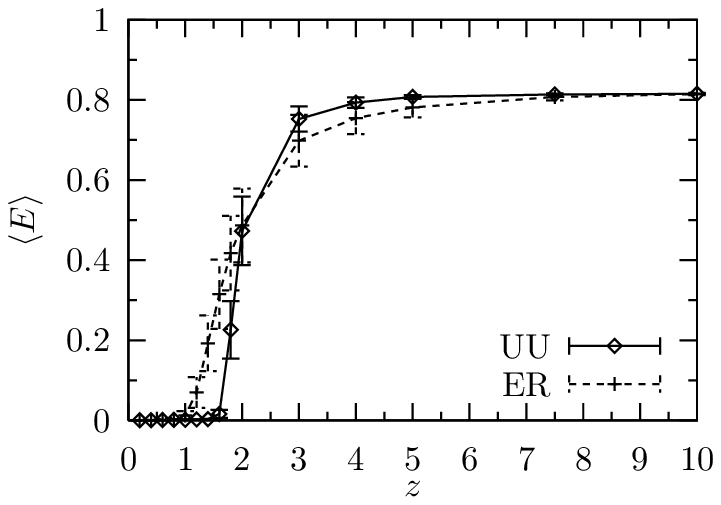}
  \hspace{0pt plus 1fil}
  \\
  \centerline{(a) \hspace{6.5cm} (b) } \\ \\
  \hspace*{0.6cm plus 1fil}
  \includegraphics[width=0.6\columnwidth]{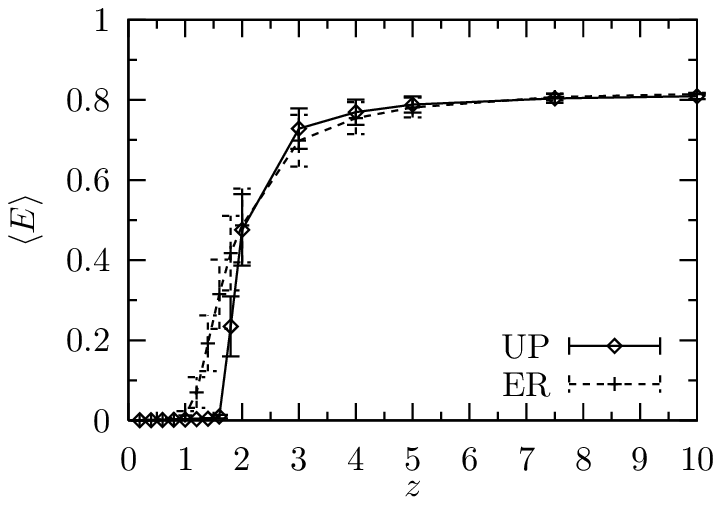} 
  \hspace{0pt plus 1fil}
  \includegraphics[width=0.6\columnwidth]{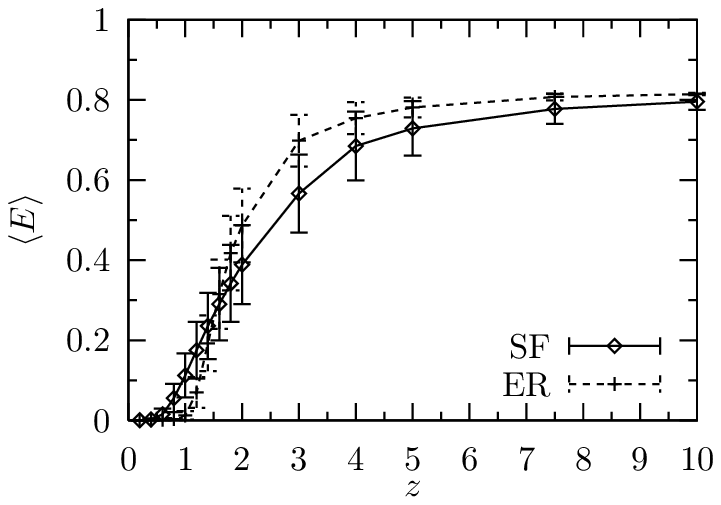}
  \hspace{0pt plus 1fil}
  \\
  \centerline{(c) \hspace{6.5cm} (d) }
 
  \caption{Average parallel efficiency for $n=1000$, $M=5000$,
    $L=100$, for the PA (a), UU (b), UP (c) and SF (d) models,
    compared with the ER model.}

  \label{fig:eff}

\end{figure*}

Figs.~\ref{fig:eff}(a--d) present the average parallel efficiency
as a function the $z$ for $M=5000$ and $L=100$. As can be easily seen by
comparing Figs.~\ref{fig:cs}(a--d) and \ref{fig:eff}(a--d), the
average parallel efficiency tends to closely reflect the normalized
average cluster size for all considered models. 
 
This is further substantiated by a comparison of the left and right
graphs of Figs.~\ref{fig:hcs}(a--e). The right graphs show the
cumulative probability distribution $P(E)$ of a randomly selected node
having efficiency greater or equal to $E$ when chosen as master node.
It can be seen that $P(E)$ closely follows $P(s).$ The main differences
are that the $P(E)$ curves are smoother and the largest achievable
values of $E$ smaller than that of $s/n.$  The latter difference is a
result of the fact that a cluster with $s$ nodes cannot achieve speedup
$S$, because the master node does not compute and the communication
costs increase the execution time of each task (in comparsion to
sequential execution); the former difference is due to the fact that
different clusters with the same number of nodes $s$ have different
interconnection topology, resulting in different communication costs.

As a consequence, similar conclusions can be drawn for the efficiency
of a grid computing network as for the mean cluster sizes of the
corresponding network model. Particularly, we note a percolation
transition of the efficiency with increasing connectivity from a
regime of very small efficiency to a regime of high efficiency. This
transition is more abrupt for the ER and particularly for the UU and UP
models, and somewhat slower for the PA and SF models. The efficiency of
the models with preferential attachment is restricted due to the already
discussed higher number of isolated nodes, that contribute to a speedup
of zero to the average speedup. The insertion models need a higher
connectivity to reach the percolation point, but after that show a
higher efficiency, demonstrating the advantage of using network
resources to connect new nodes.

\begin{figure}

  \centerline{\includegraphics[width=0.6\columnwidth]{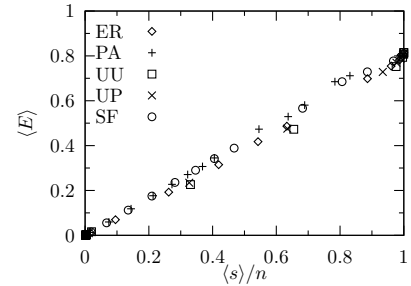}}

  \caption{Scatter plot of efficiency against cluster size, for $n
  = 1000$, $M=5000$ and $L=100$.}

  \label{fig:eXcs}

\end{figure}

The strong correlation between the cluster size and parallel
efficiency is further substantiated in Fig.~\ref{fig:eXcs}, where a
scatter plot of efficiency against normalized cluster size is shown,
that demonstrates the almost linear correlation between efficiency and
normalized cluster sizes.  Another interesting result observed from
this figure is that the PA and SF networks tend to provide slightly
better efficiencies than the other models.  In other words, if clusters
of similar size are considered, the presence of hubs that
``short-circuit'' the distances tends to enhance the speed-up of the
computations in the networks.

\begin{figure*}

  \hspace{0.6cm plus 1fil}
  \includegraphics[width=0.6\columnwidth]{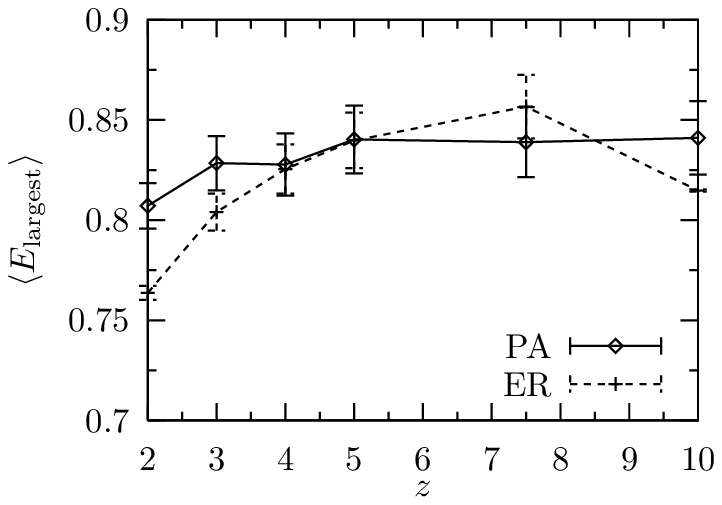}
  \hspace{0pt plus 1fil}
  \includegraphics[width=0.6\columnwidth]{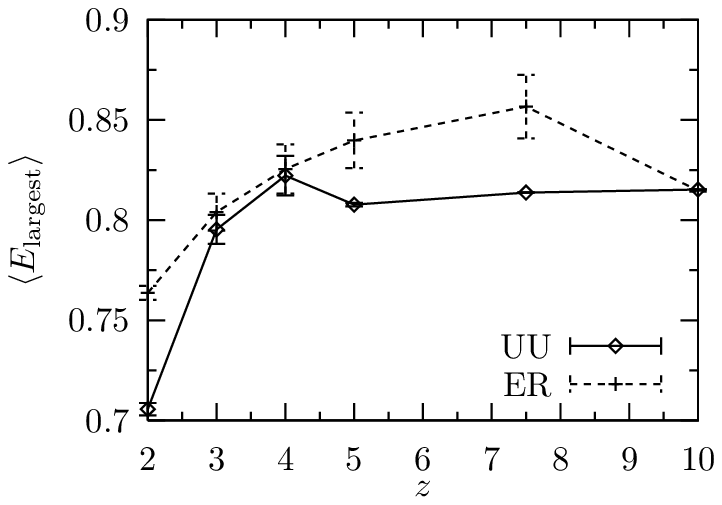}
  \hspace{0pt plus 1fil}
  \\
  \centerline{(a) \hspace{6.5cm} (b)}\\ \\
  \hspace*{0.6cm plus 1fil}
  \includegraphics[width=0.6\columnwidth]{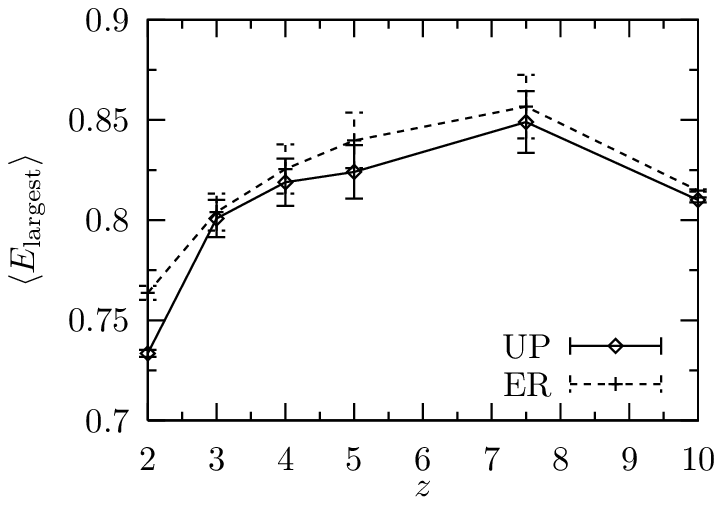}
  \hspace{0pt plus 1fil}
  \includegraphics[width=0.6\columnwidth]{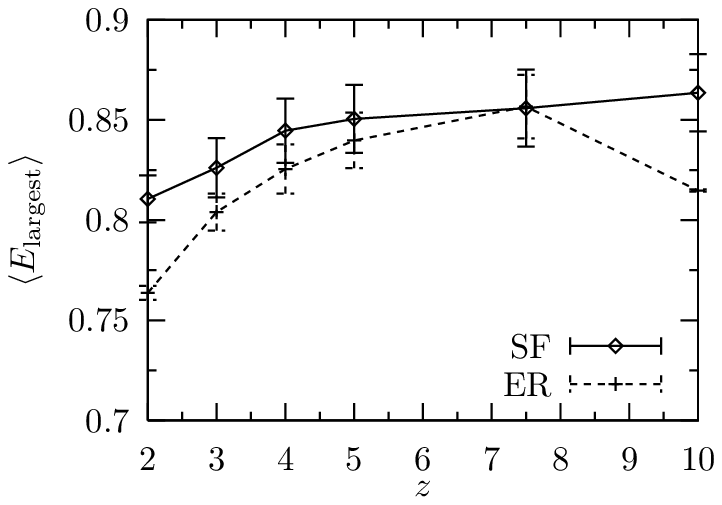}
  \hspace{0pt plus 1fil}
  \\
  \centerline{(c) \hspace{6.5cm} (d)}

  \caption{Efficiency of the largest clusters
    $\left<E_\mathrm{largest}\right>$, for the PA
    (a), UU (b), UP (c) and SF (d) models, compared with the ER model.
    In all cases $M=5000$ and $L=100$ and the size of the networks was
    chosen to have a largest cluster of about $1000$ nodes.}

  \label{fig:giant}

\end{figure*}

In order to further investigate such a possible effect of the
intrinsic connectivity properties of the considered models over the
respective performance, the largest clusters of the four models were
considered in isolation in the grid simulations. That is, networks
with the same connectivity were generated and only their largest
clusters where considered. Special care was taken so as to obtain such
connected clusters with equivalent number of nodes (about 1000 nodes).
The results are shown in Figure~\ref{fig:giant}(a--d), where the
efficiency of the parallel execution on the largest clusters, $\left<
E_\mathrm{largest}\right>$, are plotted. As we are considering here only
the largest clusters, the efficiency is computed with respect to the
number of nodes of the cluster, and not of the whole network. The
results indicate a definite tendency of the SF model, and to a lesser
extent of the PA model, to outperform the others. Such an effect is
possibly a consequence of the shorter average lengths usually observed
for this model ---see Figure~\ref{fig:l}(a--d)--- and the presence
of \emph{hubs} which act as message distribution nodes.

The results obtained for the regular topologies (hypercube and 2D and
3D torus) for $n= 1000$ are shown in Table~\ref{tab:reg}.
Interestingly, the 2D and 3D torus regular architectures led to rather
low efficiencies, despite their relatively large node degrees,
reflecting the respective large average distances, due to the absence
of shortcut links in the regular structure of these networks. The
hypercube topology allowed efficiencies comparable to the maximum
obtained for the network models, but at the expense of almost 10
connections per node, which implies a high network cost.

\section{Conclusions}

While the scale-free and preferential attachment models allowed
better efficiency considering only the largest cluster, the
Erd\H{o}s-R\'{e}nyi model tended to provide better average speedup when
all clusters where condered, as a consequence of the smaller number of
isolated clusters implied by this type of network. The insertion
models resulted in even better efficiencies, due to the inclusion of
more nodes in the largest cluster after percolation. The random models
had better efficiencies than the regular ones, due to the implied
smaller average distances. 

The results show that a network is of little use for grid computing
before the percolation point is reached, that is, for values of $z$
before the formation of a cluster spanning most of the nodes, because of
the very small resulting efficiencies. In other words, the percolation
of the network used as a grid computing resource is of fundamental
importance to the utility of the grid. Although the Internet already
connects a very large number of computers, the use of these computers
for grid computing is subject mainly to two limitations: a consent from
the part of their owner and the installation of a grid computing
software on them. It is therefore appropriate to consider the grid
network as distinct from the Internet, and two computers in the Internet
as connected in the grid network only if their owners have given
permission to use \emph{and} installed an appropriate set of protocols
and compatible software. In this aspect, the obtained results motivate
the grid community to achieve convergence in protocols and software, as
the presence of many incompatible software platforms represent the
presence of unconnected clusters of nodes in the network. This
conclusion is related with the efficiency of the network as a whole for
grid computing; for users in isolated clusters, the execution of an
application in these clusters can be nevertheless of interest, if enough
speedup is achieved.

Future work should extend the analysis to other types of distributed
systems and applications. Network models closer to some real Internet
characteristics, as in \cite{Tadic01}, should be considered. The
inclusion of measures to quantify the load of intermediate nodes
in packet transmission \cite{Goh01}, like betweenness centrality, can
improve the results, altough centrality is non-trivially related to
traffic flow if congestion is considered \cite{Holme03}. If the
assumption of communication through shortest paths (that requires global
knowledge) is relaxed, the use of local search algorithms
\cite{Adamic02} will result in stronger dependence of efficiency with
the network structure. A further refinement in this direction is to
consider queuing of packets on the nodes and congestion. Recent works
\cite{Tadic04,Tadic05} have shown that transmission times are in this
case not directly related to shortest distances, but have a much richer
behavior, depending also on the total communication load carried by the
network. 

Another important generalization is to consider complex networks
presenting links with different communication speeds. This can be
modeled using weighted networks, in which the weight of the links
reflect the bandwidth or inverse latency of the interconnecting
links. Also the nodes can display different processing powers. These
generalizations make the network models closer to real interconnection
networks.

The generalization of the parallel application model to include
communications between the tasks is of interest, expanding the classes
of applications modeled and due to the importance of the network
topology to the efficiency of these communications, resulting in an
interesting interplay between network and application characteristics.
 
We conjecture that even with the generalization of the network, routing
and application models, as suggested above, the efficiency will remain
strongly related with cluster size, altough the correlation with
shortest distance may be reduced, and other network features may
increase their importance, resulting in stronger influence of the
network model used. 

\begin{acknowledgement}
 
  L. da F. Costa is grateful to FAPESP (processes 99/12765-2 and
  96/05497-3) and CNPq for financial support. G. Travieso was
  financially supported by FAPESP grant 98/14681-8.
 
\end{acknowledgement}

%\clearpage
 
\bibliographystyle{unsrt}
\bibliography{grid}
 
\end{document}